\documentclass[%
 reprint,
 amsmath,amssymb,
 longbibliography,
 prb,
,superscriptaddress]{revtex4-2}

\usepackage[ruled]{algorithm2e}
\usepackage[noend]{algpseudocode}
\usepackage{type1cm} 
\usepackage{graphicx}
\usepackage{dcolumn}
\usepackage{bm}
\DeclareMathAlphabet{\mathpzc}{OT1}{pzc}{m}{it}
\usepackage{color}
\usepackage[usenames,dvipsnames,svgnames,table]{xcolor}
\usepackage{hyperref}

\hypersetup{pdfpagemode=FullScreen,colorlinks=true,linkcolor=NavyBlue,citecolor=BrickRed}

\begin{document}


\title{Non-Fermi liquid and antiferromagnetic correlations with hole doping in the bilayer two-orbital Hubbard model of La$_3$Ni$_2$O$_7$ at zero temperature}

\author{Yin Chen}
\thanks{These authors contributed equally to this work.}
\affiliation{Department of Physics, Renmin University of China, Beijing 100872, China}
\affiliation{Key Laboratory of Quantum State Construction and Manipulation (Ministry of Education), Renmin University of China, Beijing 100872, China}

\author{Yi-Heng Tian}
\thanks{These authors contributed equally to this work.}
\affiliation{Department of Physics, Renmin University of China, Beijing 100872, China}
\affiliation{Key Laboratory of Quantum State Construction and Manipulation (Ministry of Education), Renmin University of China, Beijing 100872, China}

\author{Jia-Ming Wang}
\thanks{These authors contributed equally to this work.}
\affiliation{Department of Physics, Renmin University of China, Beijing 100872, China}
\affiliation{Key Laboratory of Quantum State Construction and Manipulation (Ministry of Education), Renmin University of China, Beijing 100872, China}

\author{Rong-Qiang He}\email{rqhe@ruc.edu.cn}
\affiliation{Department of Physics, Renmin University of China, Beijing 100872, China}
\affiliation{Key Laboratory of Quantum State Construction and Manipulation (Ministry of Education), Renmin University of China, Beijing 100872, China}

\author{Zhong-Yi Lu}\email{zlu@ruc.edu.cn}
\affiliation{Department of Physics, Renmin University of China, Beijing 100872, China}
\affiliation{Key Laboratory of Quantum State Construction and Manipulation (Ministry of Education), Renmin University of China, Beijing 100872, China}
\affiliation{Hefei National Laboratory, Hefei 230088, China}

\date{\today}

\begin{abstract}
  High-$T_c$ superconductivity (SC) was recently found in the bilayer material La$_3$Ni$_2$O$_7$ (La327) under high pressures. We study the bilayer two-orbital Hubbard model derived from the band structure of the La327. The model is solved by cluster dynamical mean-field theory (CDMFT) with natural orbitals renormalization group (NORG) as impurity solver at zero temperature, considering only normal states. With hole doping, we have observed sequentially the Mott insulator (Mott), pseudogap (PG), non-Fermi liquid (NFL), and Fermi liquid (FL) phases, with quantum correlations decreasing. The ground state of the La327 is in the NFL phase with Hund spin correlation, which transmits the Ni-$3d_{z^2}$ ($z$) orbital inter-layer AFM correlation to the Ni-$3d_{x^2-y^2}$ orbitals. When the $\sigma$-bonding state of the $z$ orbitals ($z+$) is no longer fully filled, the inter-layer antiferromagnetic (AFM) correlations weaken rapidly. At low pressures, the fully filled $z+$ band supports a strong inter-layer AFM correlations, potentially favoring short-range spin density wave (SDW) and suppressing SC. Hole doping at low pressures may achieve a similar effect to high pressures, under which the $z+$ band intersects with the Fermi level, and consequently the spin correlations weaken remarkably, potentially suppressing the possible short-range SDW and favoring SC. 
\end{abstract}


\maketitle



\section{Introduction}

Recently, superconductivity (SC) was discovered in Ruddlesden-Popper bilayer perovskite nickelate La$_3$Ni$_2$O$_7$ (La327) with a maximum $T_c$ of about $80$ K under pressure \cite{christiansson2023Correlated} and quickly attracted a lot of attention \cite{shilenko2023Correlated,liu2023Electronic,wu2023Charge,chen2023Critical,zhang2023trends,huang2023impurity,yang2023orbital,ryee2023critical,chen2023orbital,wang2024bhtsch,Sun2023Nature,lechermann2023Electronic,sakakibara2023Possible,cao2023Flat}. In the La327, the two $e_g$ orbitals of Ni are partially filled and strongly correlated, namely the Ni-$3d_{z^2}$ (denoted as $z$) and Ni-$3d_{x^2-y^2}$ (denoted as $x$) orbitals, while the three $t_{2g}$ orbitals are fully filled. Bridged by the inner apical oxygen $p_z$ orbital in the middle of the bilayer, the $z$ orbitals of the two layers have a large effective hopping and form a low-energy $\sigma$-bonding state and a high-energy anti-bonding state. The interplay between the two $e_g$ orbitals with the bilayer structure may lead to new electron correlation behaviors. 

Previous studies using density functional theory plus dynamical mean-field theory (DFT + DMFT) have indicated that the La327 exhibits Hund correlation \cite{ouyang2024hund,Blaha-JCP152,Haule-PRB81,Georges1996rmp}. Several experiments \cite{zhang2023high,craco2024strange} have shown that the La327 features a linear resistivity behavior in the high-temperature normal state, implying a strange metal. Recently, Wang et al. \cite{wang2024non} found that the Hund spin correlation may accounts for the linear resistivity. 

Several experiments have detected spin density waves (SDWs) \cite{chen2024emela327,kh2024pisdw,ck2024La327sdw,dan2024sdwla327} at ambient pressure and low temperatures, indicating a significant interaction between magnetic and electronic properties. Since no long-range magnetic order was detected in neutron scattering measurements down to $10$ K, it is probable that the SDW is either short-range or involves only small magnetic moments \cite{xie2024,ndsl}. It is believed that magnetic fluctuations play a crucial role for the pairing mechanism of SC. Some model studies \cite{yang2023Possible,gu2023Effective,luo2023high,liu2023The} associate the $s_{\pm}$ wave pairing mechanism with inter-layer magnetic interactions between $z$-orbital spins. 

Our recent model study \cite{tian2024correlation} reports that the $z$ orbital forms an $s_{\pm}$-wave superconducting state due to strong inter-layer hopping and antiferromagnetic (AFM) correlation, but its superconducting order parameter is small; while the $x$ orbital forms a strong $s_{\pm}$-wave superconducting state by sharing the inter-layer AFM correlation of the $z$ orbitals. Suppressing the local inter-orbital spin coupling (LIOSC) \cite{tian2024correlation}, which accounts for the Hund spin correlation between the two $e_g$ orbitals of Ni, causes the Mott state to disappear and significantly suppresses SC, indicating that the LIOSC is crucial for electron correlation and SC \cite{tian2024correlation}. Usually, hole doping is a key role in manipulating SC. It is worth studying whether the $x$ orbitals can still share the inter-layer AFM correlation of the $z$ orbitals through Hund's coupling with hole doping. 

In this work, the bilayer two-orbital model of the La327 is calculated using the cluster dynamical mean-field theory (CDMFT) \cite{Georges1996rmp,Maier2005RMP} method at zero temperature, considering only the normal state. With hole doping, we have observed Mott, pseudogap (PG), non-Fermi liquid (NFL), and Fermi liquid (FL) phases sequentially, and quantum correlations decrease accordingly. The Hund spin correlation allows the $z$-orbital inter-layer AFM correlation to efficiently transmit to the $x$ orbital in the Mott, PG, and NFL phases, but this transmission fails in the FL phase. Suppressing the LIOSC eliminates the Hund spin correlation and the $x$-orbital inter-layer AFM correlation, and transitions the system from a Mott phase to a FL phase, thereby highlighting the Hund spin correlation's role in enhancing electron correlation.

Moreover, we find that the ground state of the La327 is in the NFL phase. Electron doping to fully fill the $\sigma$-bonding state of the $z$ orbitals ($z+$) can significantly enhance the inter-layer AFM correlations. At low pressures, the fully filled $z+$ band leads to strong inter-layer AFM correlations, which may promote a short-range SDW and inhibit the formation of SC order. At high pressures, the $z+$ band intersects with the Fermi level, the inter-layer AFM correlations weaken rapidly, potentially diminishing the short-range SDW and encouraging the development of SC order.


\section{Model and method}

The bilayer two-orbital Hubbard model for the La327 is employed \cite{Luo2023arXiv,tian2024correlation}. Each site contains both the two eg orbitals. The Hamiltonian is
\begin{equation}
\begin{aligned}
H&=H_0+H_I, \\
H_0&=\sum_{\rm{k} \sigma} \Psi_{\rm{k} \sigma}^{\dagger}  H(\rm{k}) \Psi_{\rm{k} \sigma}, \\
H_I&=U\sum_{il\alpha} (n_{il\alpha\uparrow}-\frac{1}{2})(n_{il\alpha\downarrow}-\frac{1}{2}) \\
&+U^{\prime}_{\rm{a}} \sum_{il\sigma}(n_{ilx\sigma}-\frac{1}{2})(n_{ilz\bar\sigma}-\frac{1}{2}) \\
&+U^{\prime}_{\rm{p}} \sum_{il\sigma}(n_{ilx\sigma}-\frac{1}{2})(n_{ilz\sigma}-\frac{1}{2}),
\end{aligned}\label{eq:h}
\end{equation}
where $\Psi_{\sigma} = (d_{Ax\sigma}, d_{Az\sigma}, d_{Bx\sigma}, d_{Bz\sigma})^{T}$ and $d_{s\sigma}$ denotes the annihilation operator for an electron with spin $\sigma$ in the states $s = Ax, Az, Bx, Bz$. Here, $A$ and $B$ label the top and bottom layers, respectively. The parameters $U^{\prime}_{\rm{a}}$ and $U^{\prime}_{\rm{p}}$ represent the inter-orbital repulsion strength for anti-parallel and parallel spins, respectively. $J$ is the Hund's coupling. The model parameters are the same as those set in Ref.~\cite{tian2024correlation}. 

Due to the mirror symmetry of this model, it is convenient to transform to the bonding ($+$) and anti-bonding ($-$) states for both $e_g$ orbitals. We define $\Psi_{\pm \rm{k}\sigma} = (c_{x \pm \rm{k}\sigma}, c_{z \pm \rm{k}\sigma})^T$ with $c_{\alpha\pm \rm{k}\sigma} =  d_{\alpha \rm{k}A\sigma} \pm d_{\alpha \rm{k}B\sigma}$. This transformation renders $H$ block-diagonal. The bonding states of the $x$ and $z$ orbitals are labeled as $x+$ and $z+$, respectively, and the anti-bonding states are labeled as $x-$ and $z-$. 

The model is calculated using CDMFT \cite{Georges1996rmp,Maier2005RMP}, with natural orbitals renormalization group (NORG) \cite{He2014prb,He2015prb,Wang2022solving} as the impurity solver. In this approach, the impurity cluster of the quantum impurity model contains two sites in a unit cell, fully considering the inter-layer quantum correlations. The detailed discussion on the convergence of CDMFT is provided in the Appendix. 


\section{Results}

\subsection{Ground state phase diagram}

\begin{figure}[t!]
  \includegraphics[width=8.6cm]{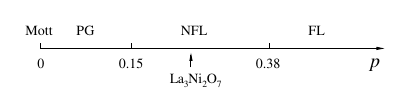}
  \caption{Ground state phase diagram with respect to hole doping $p$. Mott, PG, NFL, and FL phases are observed sequentially as $p$ increases ($p$ at the boundaries of these phases is $0, 0.15, 0.38$ corresponding to $\mu = 0.1, -1.1, -2.1$). The ground state of the La327 is a NFL state because the nominal occupancy number of Ni is $7.5$, which corresponds to 25\% hole doping (correspondingly, $\mu = -1.59$, $n_x = 0.59$, $n_z = 0.91$).}\label{fig:gs-phase-graph}
\end{figure}

\begin{figure}[b!]
  \includegraphics[width=8.6cm]{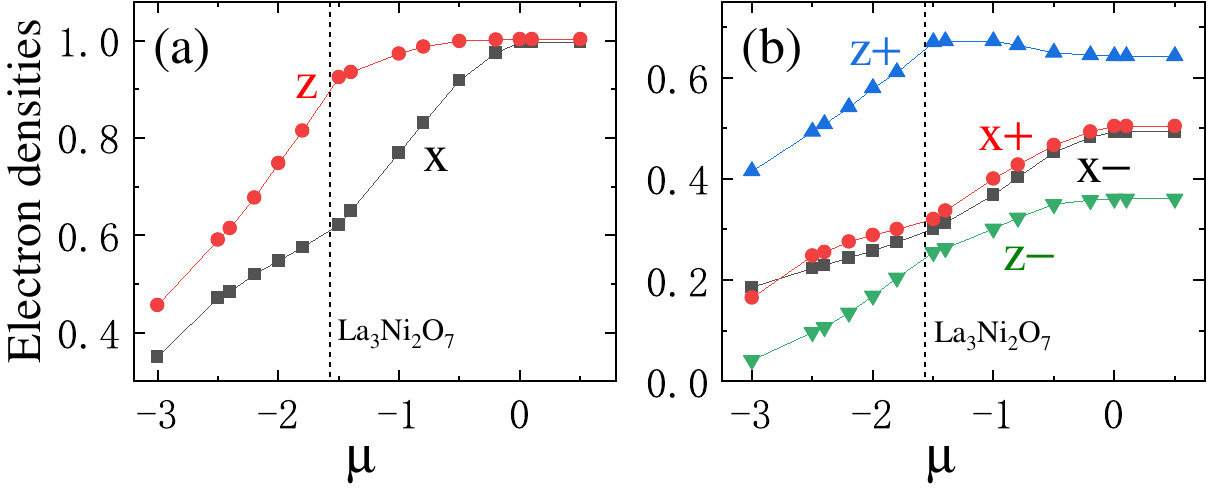}
  \caption{(a) Electron densities of the $x, z$ orbitals with respect to $\mu$. Both the $x$ and $z$ orbitals are half filling when $\mu > 0.1$. (b) Electron densities of the $x+, x-, z+, z-$ orbitals with respect to $\mu$. As $\mu$ decreases from $0.1$ to $-1.5$, the $n_{z+}$ anomalously increases, implying that the $z$ orbitals are strongly correlated. The upper boundary of the $z+$ band is at $\mu = -1.5$. The Fermi energy of the La327 is approximately $0.09$ eV below the upper boundary of the $z+$ band, consistent with previous findings in the literature \cite{ouyang2024hund}. The ground state of the La327 is a NFL state, which corresponds to $\mu=-1.59$.}\label{fig:n-mu}
\end{figure}

To study the electron correlation and spin correlation in the model with hole doping, we have done cluster DMFT calculations at zero temperature without considering any possible symmetry breaking states, i.e., only considering the normal state. According to the variation of the obtained ground-state self-energies with hole doping $p$, we draw a phase diagram (Fig.~\ref{fig:gs-phase-graph}), which contains four phases, including Mott insulator at half filling, PG at small doping, NFL at moderate doping, and FL at over doping ($p$ at the boundaries of these phases is $0, 0.15, 0.38$ corresponding to $\mu = 0.1, -1.1, -2.1$), respectively. 

\begin{figure}[t!]
  \includegraphics[width=8.6cm]{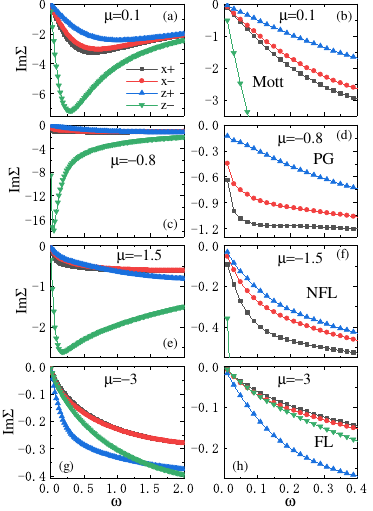}
  \caption{${\rm Im}\Sigma(i\omega_n)$ of the $x+, x-, z+, z-$ orbitals with $\mu = 0.1, -0.8, -1.5$, or $-3$ (the ground state is in the Mott, PG, or NFL phase, respectively).}\label{fig:seimp-mu0.1-0.8-1.5-3}
\end{figure}

As shown in Fig.~\ref{fig:n-mu}(a), the $x$ and $z$ orbitals simultaneously becomes half filling when $\mu > 0.1$, implying that the two orbitals take a Mott metal-insulator transition simultaneously at $\mu = 0.1$. We show the imaginary parts of orbital-dependent Matsubara self-energies (${\rm Im}\Sigma(i\omega_n)$) in different phases in Fig.~\ref{fig:seimp-mu0.1-0.8-1.5-3}. When $\mu = -0.8$, ${\rm Im}\Sigma(i\omega_n)$ of the $x+, x-, z+, z-$ orbitals have an intercept at zero frequency (Figs.~\ref{fig:seimp-mu0.1-0.8-1.5-3}(c) and (d)), indicating that the ground state is in a PG phase. When $\mu = -1.5$, ${\rm Im}\Sigma(i\omega_n)$ have no intercept at zero frequency and exhibit nonlinearity at low frequencies (Figs.~\ref{fig:seimp-mu0.1-0.8-1.5-3}(e) and (f)), indicating that the ground state is in a NFL phase. When $\mu = -3.0$, ${\rm Im}\Sigma(i\omega_n)$ have no intercept at zero frequency and are linear at low frequencies (Fig.~\ref{fig:seimp-mu0.1-0.8-1.5-3}(g) and (h)), indicating that the ground state is in a FL phase. 

Which phase does the high-pressures La$_3$Ni$_2$O$_7$ (La327) belong to when suppressing the superconducting state (namely the normal state)? It is in the NFL phase since it is 25\% hole doped (correspondingly, $\mu = -1.59$, orbital-resolved electron densities $n_x = 0.59$, $n_z = 0.91$) in our model calculation according to that the nominal occupation number of the $x$ and $z$ orbitals of the Ni atoms in La327 is $1.5$ in total. So the La327 is strongly correlated. This is consistent with the facts that the La327 has a high superconducting transition temperature and that the La327 features a linear resistivity behavior in the high-temperature normal state \cite{zhang2023high,craco2024strange}, which implies a strange metal and is theoretically studied in Ref.~\cite{wang2024non}.

\subsection{Fade of spin correlation with hole doping}

\begin{figure}[h!]
  \includegraphics[width=8.6cm]{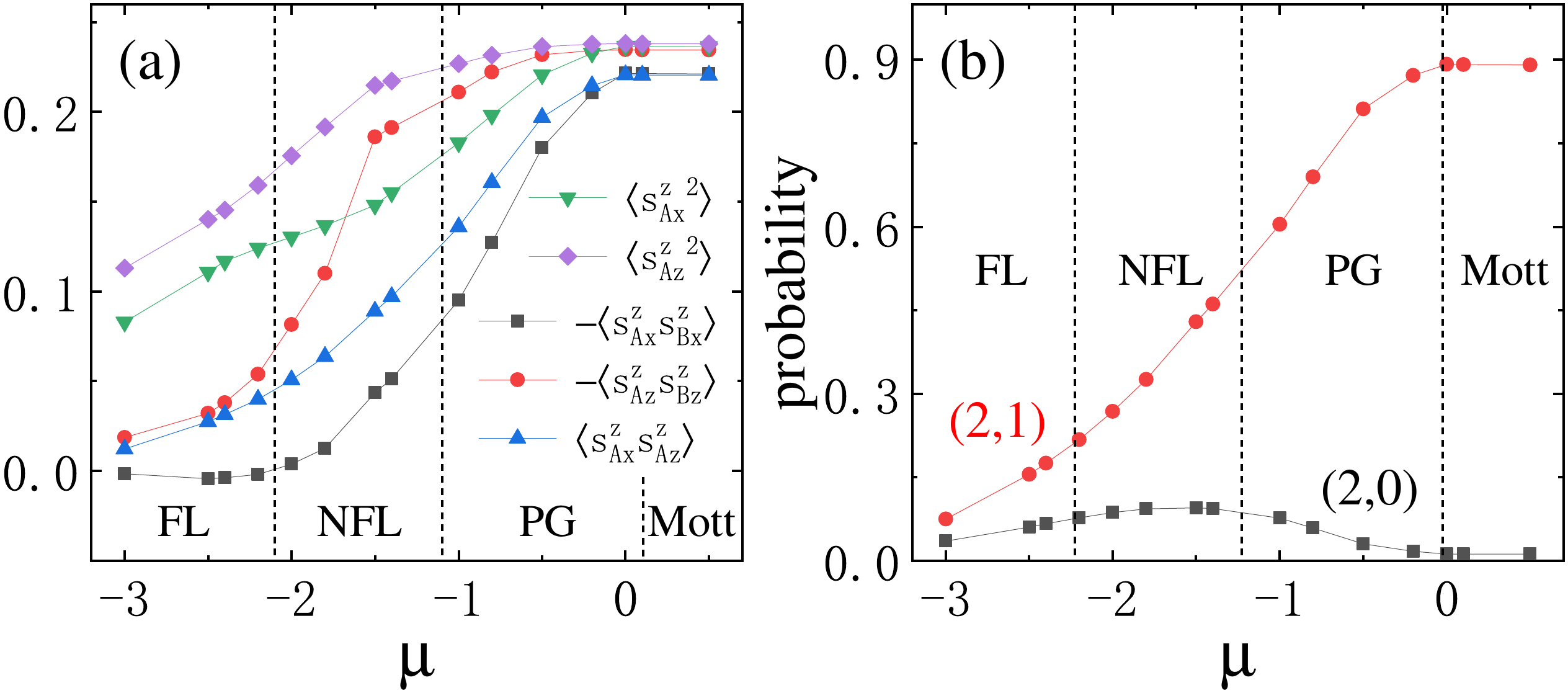}
  \caption{(a) Inter-layer AFM correlations $- \langle s^z_{Ax} s^z_{Bx}  \rangle$ and $-\langle s^z_{Az} s^z_{Bz} \rangle$, Hund spin correlation $\langle s^z_{Ax} s^z_{Az} \rangle$, and local moments $\langle (s^z_{Ax})^2 \rangle$ and $\langle (s^z_{Az})^2 \rangle$ with respect to $\mu$. (b) Probabilities of the local spin multiplets $(N_d, S_z) = (2,1)$ and $(2,0)$ with respect to $\mu$. $(N_d, S_z)$ denote the local spin multiplet in a Ni atom with $N_d$ electrons in the $x$ and $z$ orbitals and total spin $S_z$.}\label{fig:phyquan-ratio-mu}
\end{figure}

It is believed that spin fluctuation is indispensable to unconventional SC. Some studies \cite{tian2024correlation,ouyang2024hund,wang2024non} suggest that the inter-layer AFM correlations and the local Hund correlation are important for the mechanism of the proposed $s_{\pm}$-wave SC in the La327. We calculate the inter-layer AFM correlations $-\langle s^z_{Ax} s^z_{Bx} \rangle$ and $-\langle s^z_{Az} s^z_{Bz} \rangle$, the Hund spin correlation $\langle s^z_{Ax} s^z_{Az} \rangle$ of two $e_g$ orbitals of Ni, and the local moments $\langle (s^z_{Ax})^2 \rangle$ and $\langle (s^z_{Az})^2 \rangle$. The results, which show that all these quantities decrease with hole doping, are presented in Fig.~\ref{fig:phyquan-ratio-mu}(a).

The $z$-orbital inter-layer AFM correlation $-\langle s^z_{Az} s^z_{Bz} \rangle$ has an abrupt slope change at $\mu=-1.5$. The $z$-orbital local moment $\langle (s^z_{Az})^2 \rangle$ also exhibits a similar slope change. These slope changes mark a special point $\mu_{z+} = -1.5$, where the orbital-resolved electron densities (Fig.~\ref{fig:n-mu}) also change abruptly. For $\mu < \mu_{z+}$, the $z+$-orbital and $z$-orbital electron densities decrease rapidly with decreasing $\mu$. However, for $\mu > \mu_{z+}$, they remain almost unchanged, indicating that the upper boundary of the $z+$ band is at $\mu_{z+}$. When the $z+$ band is fully filled ($\mu > \mu_{z+}$), the $z$-orbital local moment and inter-layer AFM correlation decrease slowly as $\mu$ decreases; but when partially filled ($\mu < \mu_{z+}$), they rapidly decrease. This shows that the full filling of the $z+$ band, which does not mean a half filling for the $z$ orbitals (Fig.~\ref{fig:n-mu}), plays a key role for the strong magnetic correlations.

The La327 corresponds to $\mu = -1.59$ in the model (\ref{eq:h}), whose Fermi level is $0.09$ eV below the upper boundary of the $z+$ band, which is consistent with previous findings in the literature \cite{ouyang2024hund}. These imply that electron doping in the La327 will enhance the magnetic correlations. 

\subsection{Fade of Hund correlation with hole doping}
The Hund correlation of the two $e_g$ orbitals of the Ni atoms is crucial for the $s_{\pm}$-wave superconductivity in the La327 \cite{tian2024correlation,oh2023type,liao2023electron,qu2023bilayer,zhang2023structural}. To quantitatively describe the strength of the Hund correlation in model (\ref{eq:h}), we calculate the probability distribution of the local spin multiplets (LSMs) defined by the total occupancy number $N_d$ and the $z$ component $S_z$ of the total spin for a Ni site. The LSM $(N_d, S_z) = (2,1)$ is the only high-spin state among all possible LSMs. The probabilities of the LSMs $(2,1)$ and $(2,0)$ with respect to $\mu$ are exhibited in Fig.~\ref{fig:phyquan-ratio-mu}(b). The former is close to one while the latter is strictly suppressed in the Mott phase. With hole doping, the probability of the LSM $(2,1)$ decreases gradually, indicating that the Hund spin correlation \cite{wang2024non} weakens gradually. It dominates when $\mu > -2.1$, indicating that the Hund spin correlation persists, where the ground state is in the Mott, PG, or NFL phase. It no longer dominates when $\mu < -2.1$, indicating that the Hund spin correlation disappears, where the system enters the FL phase.

\subsection{Transmission of AFM correlation}

\begin{figure}[h!]
  \includegraphics[width=8.6cm]{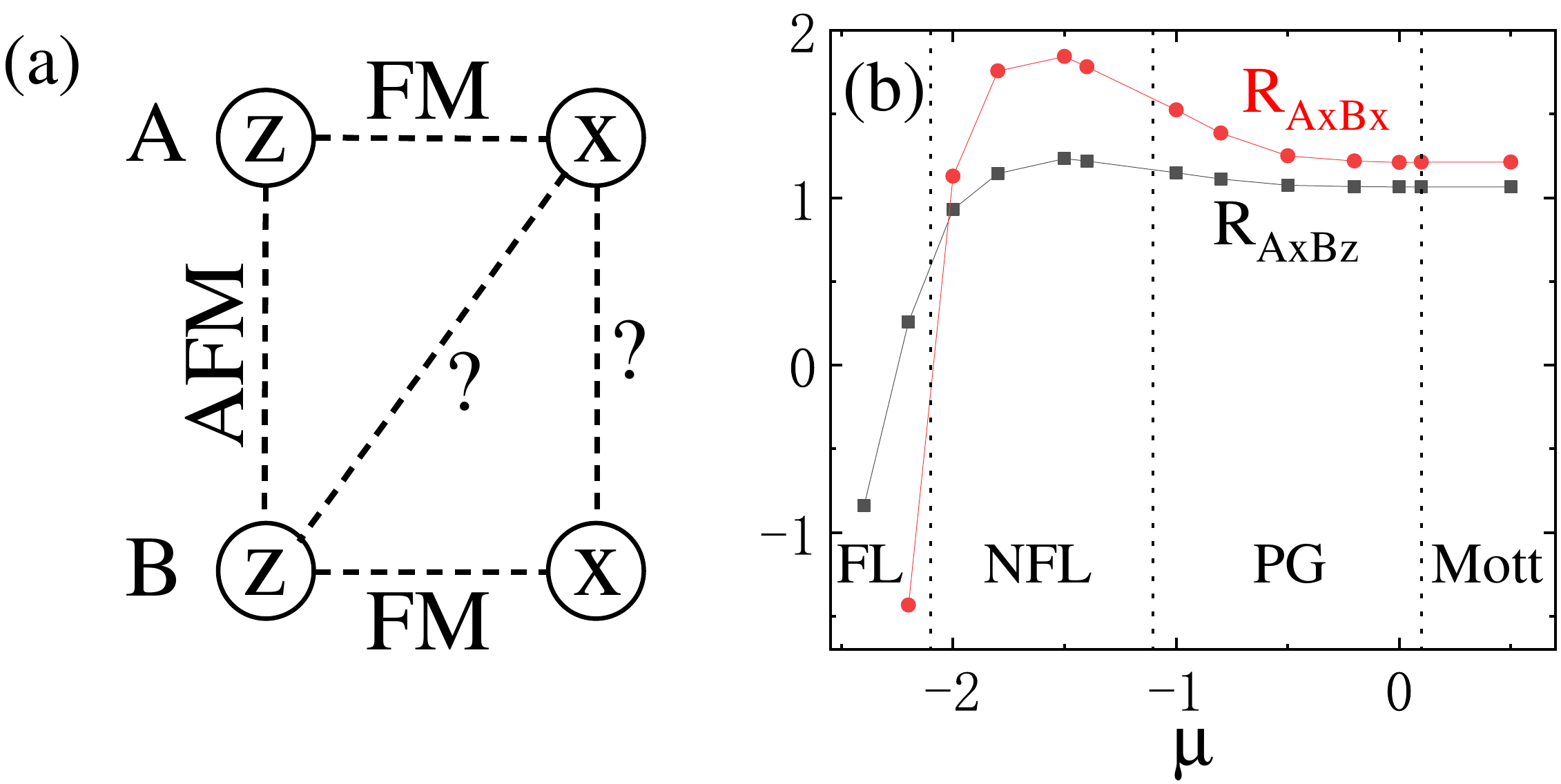}
  \caption{(a) The AFM correlation transmits from the $Az$ and $Bz$ orbitals to the $Ax$ and $Bz$ orbitals through the Hund spin correlation between the $Ax$ and $Az$ orbitals. A similar transmission occurs to the $Ax$ and $Bx$ orbitals, incorporating the additional Hund correlation between the $Bx$ and $Bz$ orbitals. (b) AFM correlation transmission functions (\ref{eq:RAxBz_RAxBx}). When $\mu > -2.1$, $R_{AxBz}$ and $R_{AxBx}$ are around $1$, indicating that the transmission is fully. When $\mu < -2.1$, they decreases drastically as $\mu$ decrease, indicating the transmission fails quickly.}\label{fig:nsz-corrtran-mu}
\end{figure}

Prior investigations \cite{tian2024correlation,ouyang2024hund,wang2024non} have shown that the inter-layer AFM correlation of $z$ orbital can be transmitted to the $x$ orbital through the Hund's coupling, which tends to align the spins on the two $e_g$ orbitals of a Ni atom. However, as hole doping increases, the Hund spin correlation weakens. In this situation, can the inter-layer AFM correlation of the $z$ orbital still be transmitted to the $x$ orbital? To answer this question, we define and calculate the inter-layer AFM correlation transmission functions 
\begin{equation}
  \begin{aligned}
    R_{AxBz}&=\frac{\langle s^z_{Ax} s^z_{Bz} \rangle}{4 \langle s^z_{Ax} s^z_{Az} \rangle \langle s^z_{Az} s^z_{Bz} \rangle}, \\
    R_{AxBx}&=\frac{\langle s^z_{Ax} s^z_{Bx} \rangle}{16 \langle s^z_{Ax} s^z_{Az} \rangle \langle s^z_{Az} s^z_{Bz} \rangle \langle s^z_{Bz} s^z_{Bx} \rangle}
  \end{aligned}\label{eq:RAxBz_RAxBx}
\end{equation}
\footnote[2]{The factors (4 or 16) in the denominator serve as normalization factors since an electron has a spin of $s = \frac{1}{2}$}. $R_{AxBz}$ describes the transmission of the inter-layer AFM correlation between $Az$ and $Bz$ orbitals to that between $Ax$ and $Bz$ orbitals through the Hund spin correlation between $Ax$ and $Az$ orbitals (Fig.~\ref{fig:nsz-corrtran-mu}(a)). $R_{AxBx}$ describes a similar transmission to that between $Ax$ and $Bx$ orbitals, including the additional Hund spin correlation between $Bx$ and $Bz$ orbitals.

When $\mu>-2.1$ (the ground state is in the Mott, PG, or NFL phase), $R_{AxBz}$ and $R_{AxBx}$ are around $1$, indicating that the transmission is fully (Fig.~\ref{fig:nsz-corrtran-mu}(b)). As $\mu$ decreases, although the $z$-orbital inter-layer AFM correlation $-\langle s_{Az}^z s_{Bz}^z\rangle$ and the Hund spin correlation $\langle s_{Ax}^z s_{Az}^z\rangle$ decrease, the AFM correlation between the $Az$ and $Bz$ orbitals can efficiently transmits to that between $Ax$ and $Bx$ orbitals through the Hund spin correlation. In stark contrast, when $\mu<-2.1$ (the ground state is in the FL phase), $R_{AxBz}$ and $R_{AxBx}$ decrease drastically as $\mu$ decreases. $-\langle s_{Ax}^z s_{Az}^z\rangle$ approaches zero with hole doping. Neither the $z$-orbital inter-layer AFM correlation nor the Hund spin correlation of the two $e_g$ orbitals of Ni is zero, but the AFM correlation between $Az$ and $Bz$ orbitals can not transmit to that between $Ax$ and $Bx$ orbitals.

\subsection{$\Delta U'$ for strong correlation}

To further understand the effect of the Hund's coupling on the spin correlation, we perform a comparative study by suppressing its spin part. We reformulate the local inter-orbital interaction on a single site as follows \cite{tian2024correlation}: 
\begin{equation}
  \begin{aligned}
      \sum _{\sigma } U^{\prime} _{\rm{a}} n_{x\sigma } n_{z\overline{\sigma }} &+U^{\prime} _{\rm{p}} n_{x\sigma } n_{z\sigma } \\
      =\frac{U^{\prime} _{\rm{a}} +U^{\prime} _{\rm{p}}}{2} n_{x} n_{z} &-2({U^{\prime} _{\rm{a}} -U^{\prime} _{\rm{p}}}) s_{x}^{z} s_{z}^{z},
  \end{aligned}\label{eq:du}
\end{equation}
where $n_{\alpha}$ represents the electron number of the $\alpha$ orbital, and $s_{\alpha}^{z}$ denotes the $z$ component of the spin of the $\alpha$ orbital. This reformulation defines the local inter-orbital spin coupling (LIOSC) \cite{tian2024correlation} as $\Delta U^\prime = U^\prime_{\rm a} - U^\prime_{\rm p}$, which exactly accounts for the Hund spin correlation between the two $e_g$ orbitals of Ni. When suppressing the LIOSC, Ref.~\cite{tian2024correlation} shows that the Mott phase disappears and the SC is suppressed remarkably.

\begin{figure}[htbp!]
  \includegraphics[width=8.6cm]{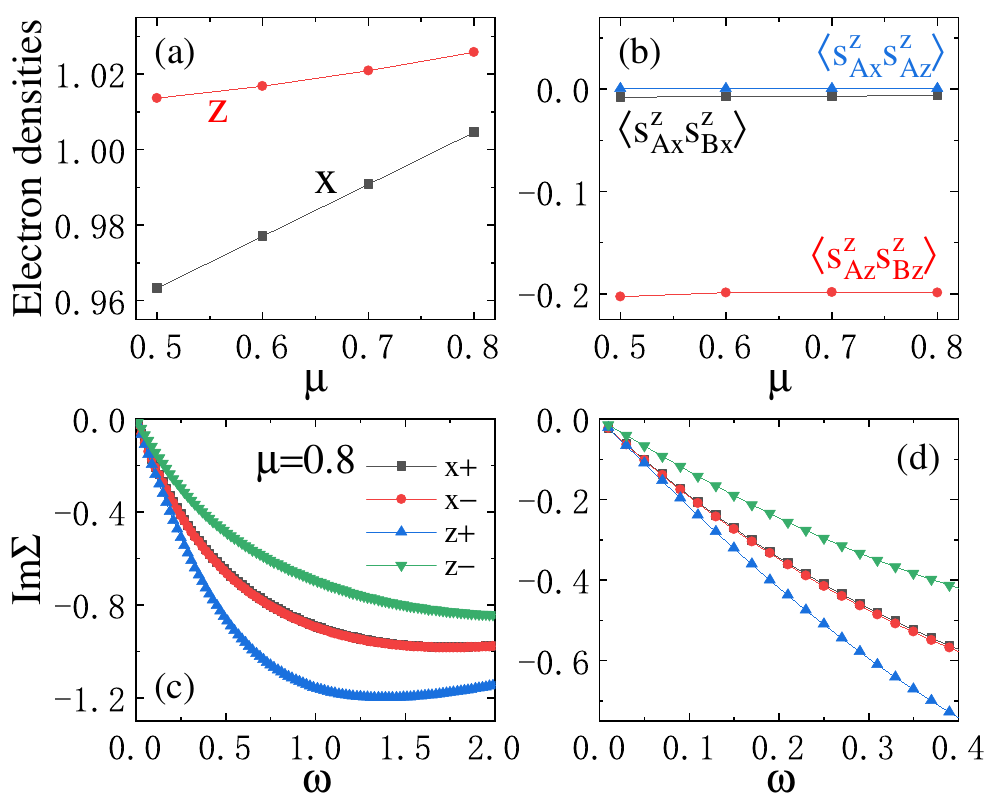}
  \caption{Calculations for $\Delta U' = 0.002J$. (a) Electron densities of the $x$ and $z$ orbitals with respect to $\mu$. Orbital-resolved electron densities $n_x$ and $n_z$ increase continuously as $\mu$ increases, without plateauing at $1$ (unit electron density), indicating absence of the Mott phase. (b) The $x$-orbital inter-layer AFM correlation $\langle s^z_{Ax} s^z_{Bx} \rangle$ and the Hund spin correlation $\langle s^z_{Ax} s^z_{Az} \rangle$ are close to zero. The $z$-orbital inter-layer AFM correlation $\langle s^z_{Az} s^z_{Bz} \rangle$ remains finite. (c) ${\rm Im}\Sigma(i\omega_n)$ of $x+, x-, z+, z-$ orbitals with $\mu = 0.8$. ${\rm Im}\Sigma(i\omega_n)$ has no intercept at zero frequency and is linear at low frequencies, indicating a FL state. (d) An enlarged view for (c) at low frequencies.}\label{fig:deltaU0.002-n-afc-seimp-mu}
\end{figure}

We do a calculation where $(U^\prime_{\rm a} + U^\prime_{\rm p})/2$ remains unchanged while $\Delta U^\prime$ is reduced from $J$ to a negligible value of $0.002J$. We find that $n_x$ and $n_z$ increase continuously as $\mu$ increases without plateauing at half filling (Fig.~\ref{fig:deltaU0.002-n-afc-seimp-mu}(a)), indicating that the Mott phase with a large Mott gap disappears. ${\rm Im}\Sigma(i\omega_n)$ has no intercept at zero frequency and is linear at low frequencies (Figs.~\ref{fig:deltaU0.002-n-afc-seimp-mu}(c) and (d)), indicating that the ground state is a FL. This shows that the LIOSC enhances pronouncedly the electron correlation.

We also find that the $x$-orbital inter-layer AFM correlation $\langle s_{Ax}^z s_{Bx}^z\rangle$ and the Hund spin correlation $\langle s_{Ax}^z s_{Az}^z\rangle$ approach zero, but the $z$-orbital inter-layer AFM correlation $\langle s^z_{Az} s^z_{Bz} \rangle$ remains finite, as shown in Fig.~\ref{fig:deltaU0.002-n-afc-seimp-mu}(b). This shows that the LISOC plays a key role in transmitting the inter-layer AFM correlation of $z$ orbitals to the $x$ orbitals through Hund's coupling. This indicates that the inter-layer AFM correlation of the $z$ orbitals cannot be transmitted to the $x$ orbitals through Hund's coupling. When the LISOC is not suppressed, the inter-layer AFM correlation of the $x$ orbital is mainly derived from that of the $z$ orbital through Hund's coupling. 


\section{Discussion about La327}

The La327 has a full filling of the $z+$ band and short-range SDW at low pressures \cite{taoxiangNSPL2024,chen2024emela327,kh2024pisdw,ck2024La327sdw,dan2024sdwla327,xie2024,ndsl}, with no SC observed. Under high pressures, the $z+$ band is slightly hole-doped, resulting in high-$T_c$ SC. Usually, SDW and SC compete with each other \cite{2019FACOSCAFMHC}.

Our model calculations show that with hole doping, the $z+$ band is not fully filled, the $z$-orbital inter-layer AFM correlation weakens remarkably. Simultaneously, the $x$-orbital inter-layer AFM correlation also decreases. Whether the $z+$ band is fully filled or not significantly affects the strength of the AFM correlation, thereby influencing the competition between the short-range SDW and the SC. For the La327, under low pressures, the $z+$ band is fully filled, and the inter-layer AFM correlation is relatively strong, which may help to stabilize the short-range SDW, thus suppressing the establishment of SC order. If the $z+$ band is doped and intersects the Fermi level, it rapidly weakens the inter-layer AFM correlations and may diminish the SC. If the $z+$ band is over doped, the AFM correlation will weaken excessively, failing to provide sufficient pairing interaction, thereby killing the SC \cite{tian2024correlation}.

\begin{figure}[t!]
  \includegraphics[width=8.6cm]{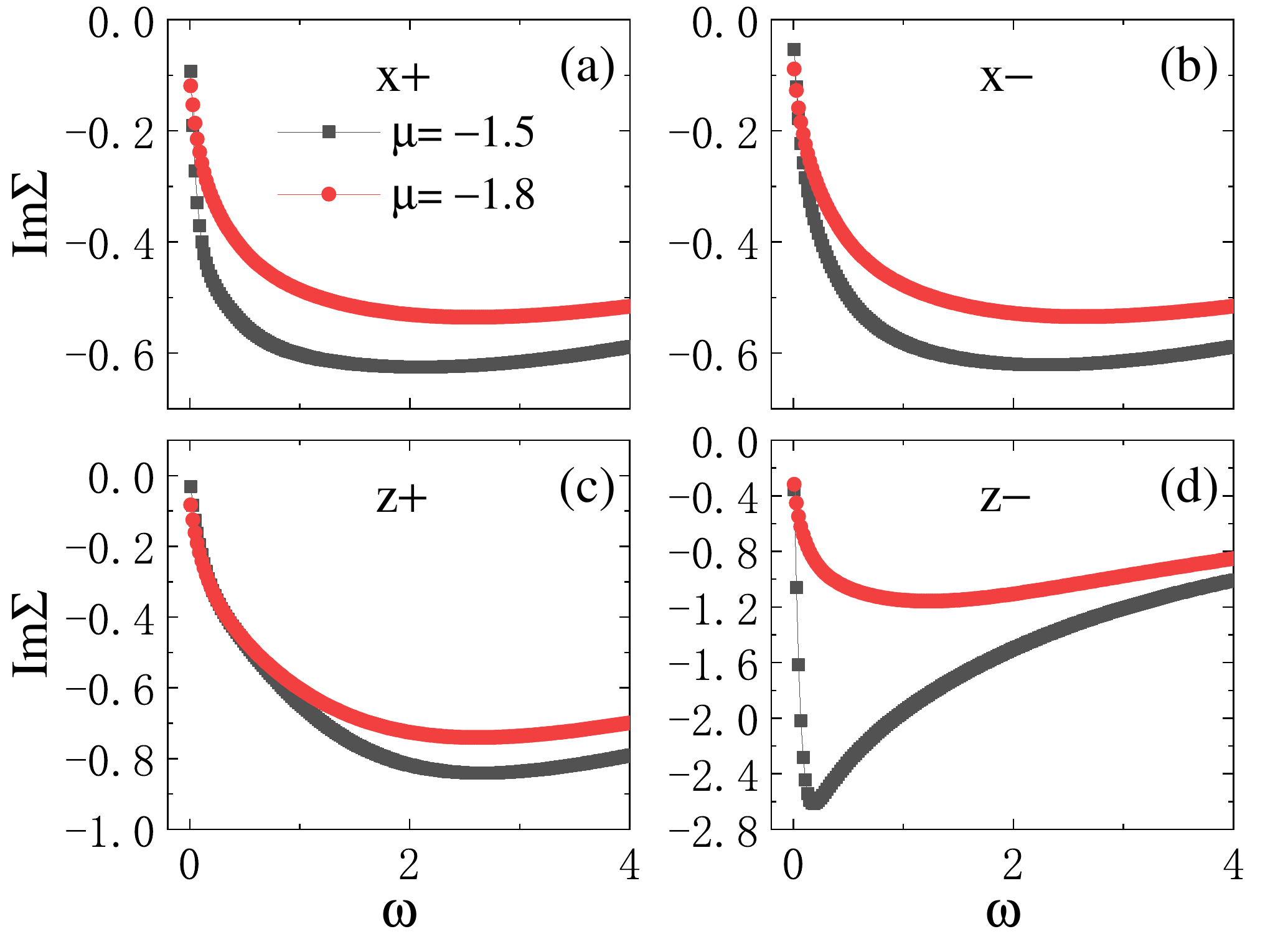}
  \caption{${\rm Im}\Sigma(i\omega_n)$ of the $x+, x-, z+, z-$ orbitals at $\mu = -1.5$ and $-1.8$.}\label{fig:mu-1.5-1.8_4sets_ImSigma}
\end{figure}

At low pressures, the $z+$ band in the La327 is fully filled, resulting in strong inter-layer AFM correlations. Meanwhile, the $z$ orbital becomes half-filled and the band is narrow, thus exhibiting strong Mott correlations. The $x$ orbital is correlated with the $z$ orbital through the Hund's coupling within the Ni, which enhances the correlation of the $x$ orbital as well. As the pressures increases, the $z+$ band intersects with the Fermi level, causing the inter-layer AFM correlations to weaken rapidly. Simultaneously, the $z$ orbital becomes less than half-filled, and the Mott correlation weakens, leading to a reduction in the correlation of the $x$ orbital as well. Consequently, significant changes in the correlated electronic structure, such as the substantial variations in the self-energies (Fig.~\ref{fig:mu-1.5-1.8_4sets_ImSigma}), are observed before and after the $z+$ band crosses the Fermi level. This is helpful for understanding the electron correlations in the La327 and, on this basis, for studying the mechanism of high-$T_c$ SC. 


\section{Summary}

\begin{figure}[htbp!]
  \includegraphics[width=8.6cm]{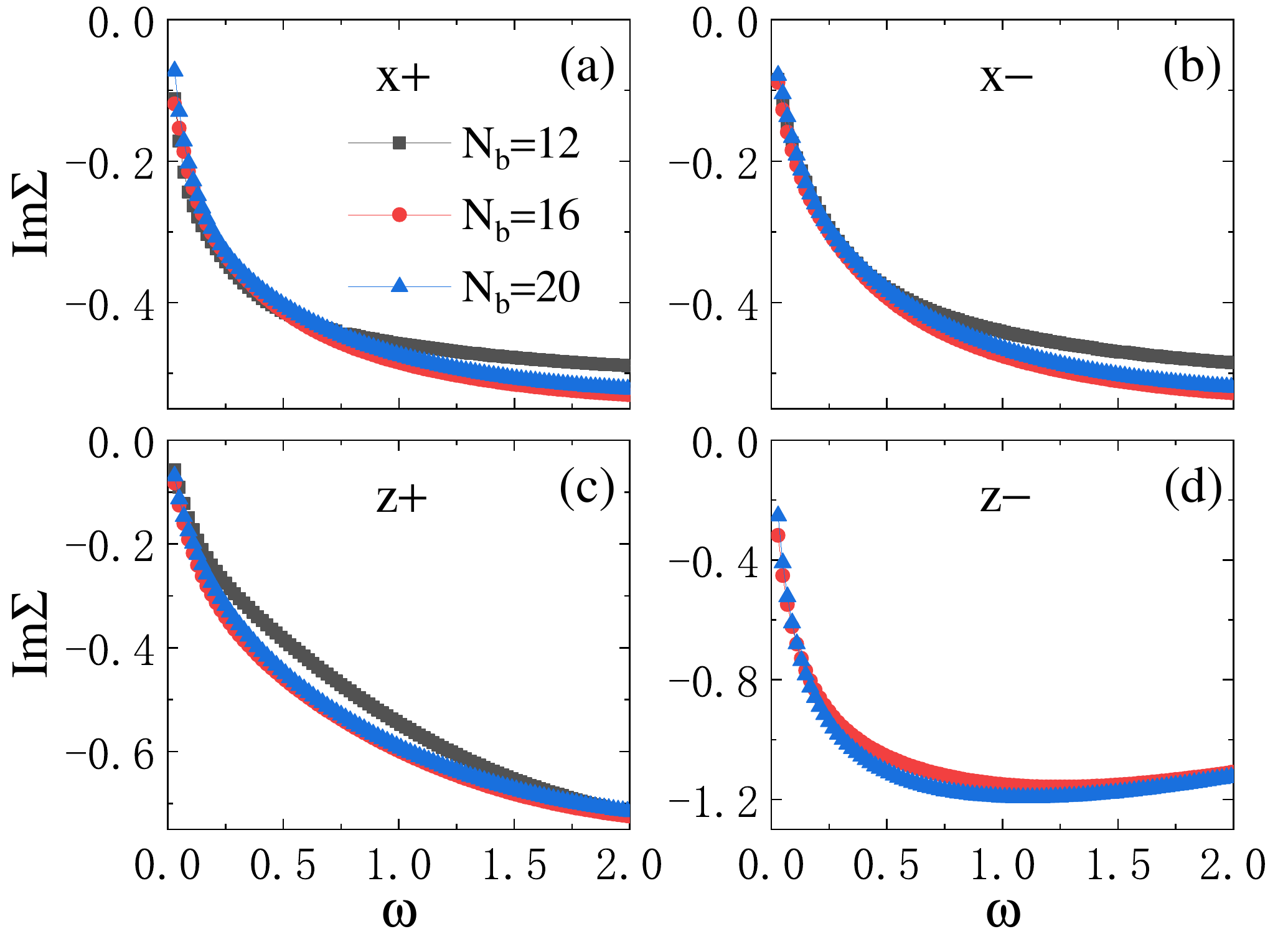}
  \caption{${\rm Im}\Sigma(i\omega_n)$ of the $x+, x- , z+, z-$ orbitals at $\mu = -1.8$ are obtained by NORG+CDMFT with $N_b = 12$, $16$, and $20$ bath sites, respectively.}\label{fig:mu-1.8_345sets_ImSigma}
\end{figure}

At zero temperature, the bilayer two-orbital model (\ref{eq:h}) of the La327 is calculated using the CDMFT method without considering any potential symmetry breaking states, i.e., considering only the normal state. We find that as increasing hole doping, the ground state sequentially transitions through the Mott, PG, NFL, and FL phases. Hole doping weakens various quantum correlations. Once the $z+$ band is no longer fully filled, the inter-layer AFM correlations will rapidly weaken. When the ground state is in the Mott, PG, or NFL phase, the Hund spin correlation persists, allowing the inter-layer AFM correlation of the $z$ orbital to transmit to the $x$ orbital. And particularly this transmission is about 100\%. But it quickly breaks down and the $x$-orbital inter-layer spin correlation vanishes when the system enters the FL phase. When suppressing the LIOSC, the Hund spin correlation and the $x$-orbital inter-layer AFM correlation disappear. And the system becomes a FL state, showing that the Hund's coupling can greatly enhance electron correlation.

\begin{figure}[t!]
  \includegraphics[width=8.6cm]{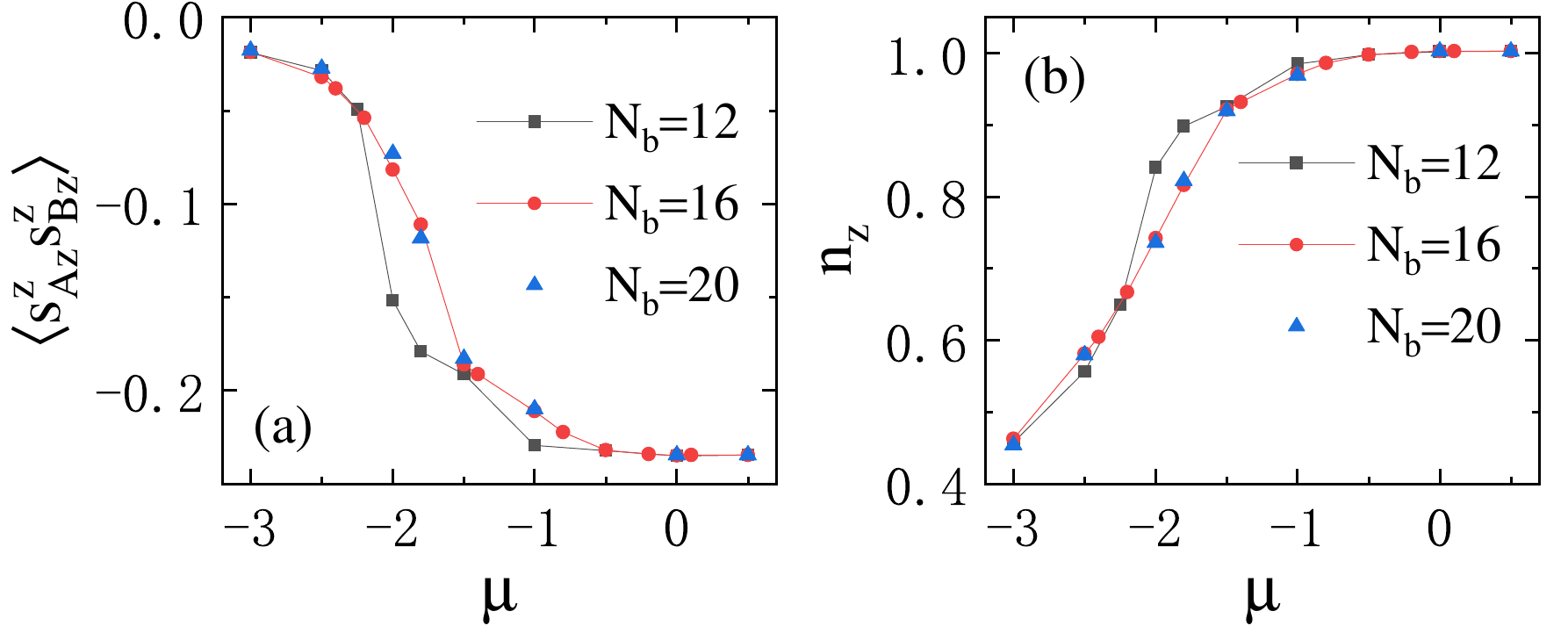}
  \caption{$z$-orbital inter-layer spin correlation $\langle s^z_{Az} s^z_{Bz} \rangle$ (a) and electron density $n_z$ (b) with respect to the chemical potential $\mu$ are obtained by NORG+CDMFT with $N_b = 12$, $16$, and $20$ bath sites, respectively.}\label{fig:345sets-Nzafmcr-mu}
\end{figure}

At low pressure, the fully filled $z+$ band in the La327 leads to large local moments and strong inter-layer AFM correlations. This may support the short-range SDW and suppress the SC order. At low pressures, hole doping may have an effect similar to that of high pressures, causing the $z+$ band to intersect with the Fermi level. This can significantly weaken spin correlations, and may hence suppress the short-range SDW and promote SC.

\begin{acknowledgments}
  This work was supported by National Natural Science Foundation of China (Grant No. 11934020). Z.Y.L. was also supported by Innovation Program for Quantum Science and Technology (Grant No. 2021ZD0302402). Computational resources were provided by Physical Laboratory of High Performance Computing in Renmin University of China. 
\end{acknowledgments}

\appendix

\section{\label{appA} Convergence with the number of bath sites in CDMFT}

The electronic bath for the impurity model of CDMFT is discretized into a number of bath sites. Bath discretization may lead to error. Ref.~\cite{Wang2022solving} shows that the error decreases exponentially as the number of bath sites increases. To determine the number of bath sites required for CDMFT to converge, we performed calculations with 12, 16, and 20 bath sites, respectively. We find that ${\rm Im}\Sigma(i\omega_n)$ for the $x+, x-, z+, z-$ orbitals (Fig.~\ref{fig:mu-1.8_345sets_ImSigma}), as well as $\langle s^z_{Az} s^z_{Bz} \rangle$ and $N_z$ (Fig.~\ref{fig:345sets-Nzafmcr-mu}), converge with 16 and 20 bath sites. We utilize 16 bath sites for the impurity model  in our CDMFT calculations. 

\bibliography{lno327afm}

\end{document}